\documentclass[review,number]{elsarticle}

\biboptions{sort&compress}

\usepackage{graphicx}

\usepackage{amssymb}
\usepackage{amsmath}
\usepackage{tabularx}

\begin{document}

\begin{frontmatter}

\title{Effect of assessment error and private information on stern-judging in indirect reciprocity}

\author[label1]{Satoshi Uchida\corref{cor1}}
\ead{s-uchida@rinri-jpn.or.jp}
\address[label1]{%
Research Division, RINRI Institute,\\
Chiyoda-ku misaki-cho 3-1-10, 101-8385 Tokyo, Japan}%
\author[label2,label3]{Tatsuya Sasaki}
\ead{tatsuya.sasaki@univie.ac.at}
\address[label2]{%
Faculty of Mathematics, University of Vienna, \\
Oskar-Morgenstern-Platz 1,1090 Vienna, Austria}
\address[label3]{Evolution and Ecology Program, International Institute for Applied Systems Analysis, \\
Schlossplatz 1, 2361 Laxenburg, Austria
}
\cortext[cor1]{Corresponding author. Tel.: +81 422 522770 Fax.: +81 422 522520}

\begin{abstract}
Stern-judging is one of the best-known assessment rules in indirect reciprocity.
 Indirect reciprocity is a fundamental mechanism for the evolution of cooperation.
It relies on mutual monitoring and assessments, 
i.e., individuals judge, following their own assessment rules, whether other individuals are ``good'' or ``bad'' according to information on their past behaviors.
Among many assessment rules, stern-judging is known to provide stable cooperation in a population,
as observed when all members in the population know all about others' behaviors (public information case) 
and when the members never commit an assessment error.
In this paper, the effect of assessment error and private information on stern-judging is investigated.
By analyzing the image matrix, which describes who is good in the eyes of whom in the population, 
we analytically show that private information and assessment error cause the collapse of stern-judging:
all individuals assess other individuals as ``good" at random with a probability of 1/2.
\end{abstract}

\end{frontmatter}

\section{Introduction}

``I help you and you help me''. Such cooperative relationships can be often found in biological systems and human societies. Cooperative behaviors are clearly important to make biological and human societies effective and smooth. However, evolutionary biologists and social scientists have long been puzzled about the origin of cooperation \cite{Pennisi05}. Recently, scientists from a variety of fields such as economics, mathematics or even physics tackle the puzzle using the power of mathematics \cite{Perc13}.

This puzzle is called the social dilemma or the free rider problem \cite{Dawes80}. This can be described as follows: (1) individuals in a society have binary choices: cooperation (help others) or defection (refuse to help others), (2) a society consisting of cooperators is more profitable than that with only defectors, (3) but, within a society, individual defectors do better than individual cooperators, since cooperators must incur a cost to help others, while defectors do not. Thus it is more advantageous for individuals to choose defection regardless of what other individuals choose, which, by natural selection or social learning, leads to a society with only defectors. The fundamental issue therefore is to explain why cooperation is so ubiquitous in the real world.

Indirect reciprocity is one of the basic mechanisms known to sustain mutual cooperation among individuals 
\cite{Alexander87,Nowak06,Sugden86,Trivers71,RandEtal13}. It is found not only in human societies \cite{Camerer06,Bolton05,Seinen01,Wedekind00,Wedekind02,VuHa10,BaRe13,YoEtal13} but also in biological systems \cite{Bshary06}.
Indirect reciprocity is also a boosting mechanism of group formation \cite{OishiEtal13} and in-group favouritism \cite{Masu12,NaMa12,FuEtal12}, another aspect of human cooperation.
This mechanism relies on the abilities of individuals to monitor the interactions of others, and to assess those interactions using moral sense even when they are not personally engaged in them \cite{Leimar01, Nowak98a, Nowak98b, Panchanathan03,SuAki08,FuEtal08, Berg11, Nakamura11, CongEtal12, TanaEtal12, SuKi13, MarCue13}.

If the action of an individual is assessed as bad in the population, or in other words, if the individual is burdened with a bad image in the society, the individual will not be given help from members of that society anymore. Thus, by indirect reciprocity, helpful actions can be channeled away from bad individuals, and directed to good individuals. Therefore, if defectors are labeled with a bad name and if a good image is assigned to just helpers, indirect reciprocity makes it beneficial to help others, even when the helpful actions incur some cost to the helpers. As a result, members in a society opt to help each other so that cooperative relationships evolve.

In assessing others' behaviors, members follow an assessment rule by which to judge the acts of others as good or bad. This can be viewed as an elementary model of a social norm of the society. There are clearly many possibilities for those rules, each of which corresponds to a moral culture of a society (i.e., what are regarded as good acts and what are bad acts in the society) \cite{Brandt04,Ohtsuki04}.

The simplest rule only takes it into account whether help is given or not and does not use any other information such as who provided the help to whom. Plausibly, a good image is assigned to individuals who give help to others and a bad image to individuals who refuse to help \cite{Nowak98a}. 

However, this simple assessment rule gives rise to an inconsistency and cannot provide a stable cooperation in a population. Using this rule, refusing to help is always assigned a bad image. Therefore, when an individual $i$ refuses to help $j$ because $j$ is a bad person, $i$ is perceived by other individuals as bad, although they would have also refused to help $j$. And those individuals who refused to help $i$ become bad again. In this way, bad images are copied and spread in the population so that in the end the population consists of bad individuals.

This example indicates that not all assessment rules lead to stable cooperation. In fact, Ohtsuki and Iwasa \cite{Ohtsuki04,Ohtsuki06} showed that, assuming binary assessments (i.e., a world where there are only good and bad assessments), only eight lead to a stable regime of mutual cooperation among all possible rules.
 These are called the leading eight.
Two of these rules are based on the so called second-order assessments which take into account whether the image of the recipient is good or bad as well as whether help is given or not. The other six are based on third-order assessments which use the additional information on whether the donor is good or bad \cite{Brandt04}.

In order to discuss the advantages and disadvantages of these rules,
Uchida and Sigmund \cite{Uchida10a} proposed a method to compare different assessment models.
They applied this method to compare the two second-order assessment rules
 and analytically showed that the sterner second-order rule called stern-judging \cite{Kandori92} wins in the sense that 
players using stern-judging are more likely to earn a higher payoff than players following the other milder rule (called standing) \cite{Sugden86}.
Other papers based on numerical simulations also argue the advantages of stern-judging \cite{Brandt04,Chalub06,Pacheco06}.

These studies, however, assume that the information is public and that no personal assessment error occurs, 
i.e., all members in a population can know about the behaviors of all others and they assess the behaviors according to their assessment rules without any personal mistake. 
As a result, all individuals always agree about an assessment of other individuals.
Daily experience, however, tells us that persons can have different information \cite{Uchida10b,Nakamura11} 
and that they can hold personal misconceptions when assessing others \cite{Takahashi06}.
Such different information and mis-perceptions can lead to a mismatch between the opinions of individuals,
and possibly cause the collapse of assessment rules that are deemed successful in the case of public information without assessment errors.
As a first step to theoretically investigating the effects of assessment errors and private information on the leading eight,
 we consider stern-judging in this paper.

In the literature, the impact of private information is often investigated numerically by using the concept of image matrix \cite{Uchida10b}. 
The image matrix describes who is good in the eyes of whom in the population.
However, its ``rigorous analysis seems to offer considerable challenges" \cite{Sigmund12}.
In this paper, we assume that only some of the members in a population can observe each game
and that the observers can personally make a mistake in assessing the donor in the game.
Under this setting, and by investigating image matrix, we analytically calculate the proportion of good individuals in a society
in which stern-judging pervades as a social norm.
We find that the proportion of good individuals is always 1/2 with the exception of the special case of public information without any assessment error.

The following sections describe the model and the methods of analysis, 
then derive the results, and discuss both the model and outcomes.

\section{The Model}
A large, well-mixed population of players is considered.
From time to time, two players are chosen at random from the population 
and they engage in a one-shot donation game:
 a coin toss decides who plays the role of potential donor and who is recipient.
The donor decides whether or not to help the recipient, at a personal cost $c$.
If the donor chooses to help, the recipient gains a benefit $b>c$;
otherwise the recipient obtains nothing.
Each individual in the population experiences such donation games many times \cite{Sigmund10}. 

In indirect reciprocity, individuals in the population have an ability to observe other individuals.
Each game is observed by a fraction $q$ of the population.
This assumption is different from that of Uchida and Sigmund \cite{Uchida10a} who assume that all individuals perfectly observe all interactions, 
and from Ohtsuki and Iwasa \cite{Ohtsuki04,Ohtsuki06}  who assume that one randomly chosen individual acts as a referee 
and whose assessment is perfectly representative of all other individuals. 
These two studies correspond to $q=1$.
In this paper, the case of private information, i.e., $q<1$ is investigated.

If an individual observes a game, the individual evaluates the action of the donor in the game.
Each individual has an assessment rule by which to judge the action of the donor. 
We assume a binary judgment: either the label ``good'' or ``bad'' is assigned to the donor.
In this paper, the assessment rule called stern-judging is considered.
Stern-judging views those as good who, in their previous game, gave help to a good recipient or refused help to a bad recipient \cite{Kandori92,Ohtsuki06}.

We assume that an assessment error occurs with a small probability.
With a probability $\mu$, an observer assigns the opposite assessment value
to the assessment value given by the stern-judging.
This kind of mis-perception occurs individually, therefore, leads to a difference in the opinions (assessments) of players in the population, even when they follow the same assessment rule.
 
 Each individual determines whether or not to help the recipient in a game according to the current image of the recipient (i.e., whether the recipient is good or bad).
If the recipient is viewed as good in the eyes of the potential donor, the recipient will be given help, or otherwise refused to help.
But the donor can commit an implementation error: 
with a certain probability $\epsilon$, the donor fails to implement an intended help. 
Following previous studies, 
an intended refusal is assumed to be always carried out \cite{Leimar01,Ohtsuki04,Ohtsuki06,Panchanathan03}.

The payoff a random recipient obtains from a random donor depends on the frequency of good individuals in the population.
Let $r$ denote the probability that a random individual positively 
assesses another random individual.
Then the expected payoff is given by $\bar\epsilon r (b-c)$ with $\bar\epsilon=1-\epsilon$.
The fully cooperative population corresponds to $r=1$ and the fully defective population to $r=0$.
In the next section, we describe how the value of $r$ is determined.

\section{Image matrix}

Let $\beta_{ij}$ denote the image of player $j$ in the eyes of player $i$.
The values $\beta_{ij}=1$ and $\beta_{ij}=0$ correspond to the situations where $i$ thinks that $j$ is good and bad respectively.
The matrix $(\beta_{ij})$ is called the image matrix \cite{Uchida10b}. 
Then the value of $r$ is the mean of all the elements of $(\beta_{ij})$.

\begin{figure}
\begin{center}
\includegraphics[scale=0.5]{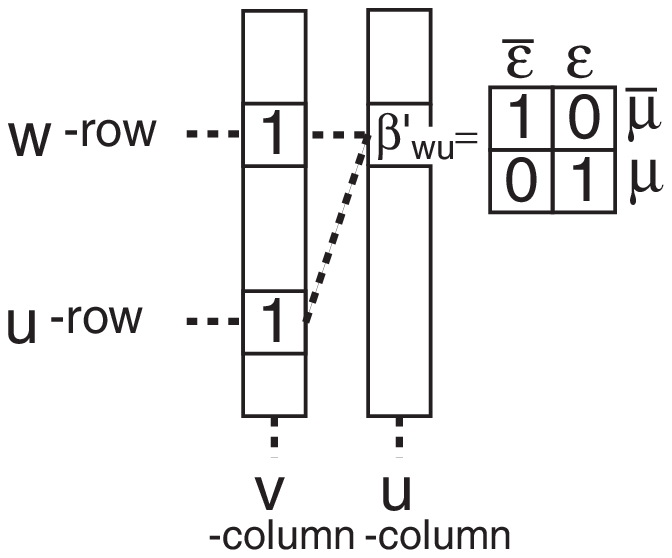}
\includegraphics[scale=0.5]{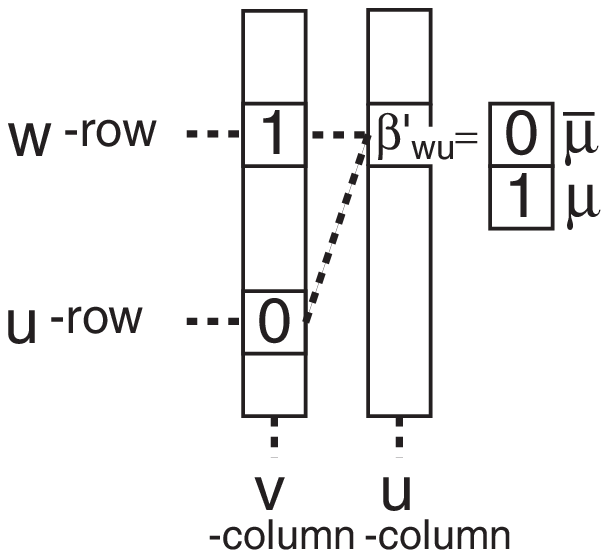}\\
 \hspace{-25mm}(a)  \hspace{35mm}(b)\\
\includegraphics[scale=0.5]{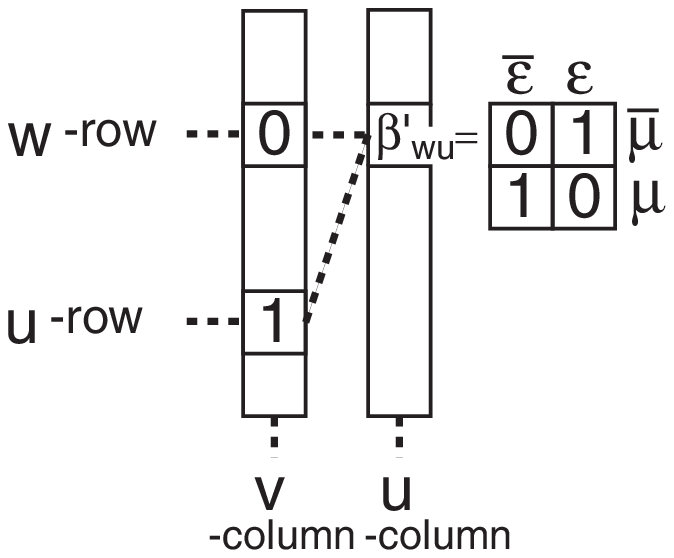}
\includegraphics[scale=0.5]{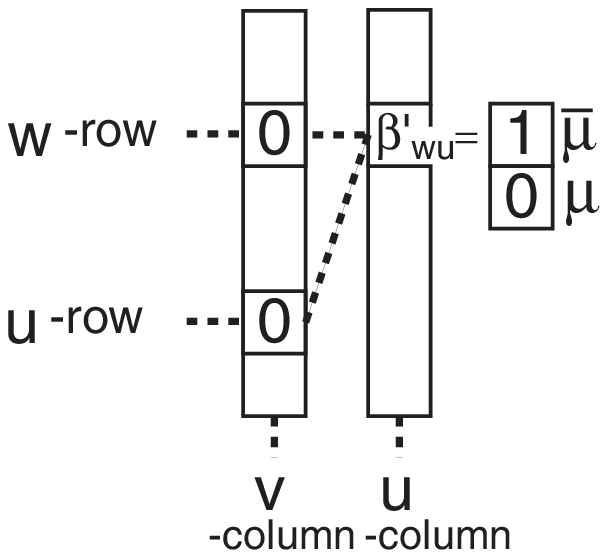}\\
 \hspace{-25mm}(c)  \hspace{35mm}(d)\\
\caption{Update of the image matrix.
The donor, the recipient and one of the observers at the game are denoted by $u$, $v$ and $w$ respectively. In the figure, $\beta'_{wu}$ denotes the new assessment of donor $u$ in the eyes of observer $w \in {\cal W}$.
The new assessment depends on the current assessment of $v$ both in the eyes of 
$u$ ($\beta_{uv}$) and $w$ ($\beta_{wv}$):
(a) $\beta_{uv}=\beta_{wv}=1$, (b) $\beta_{uv}=0, \beta_{wv}=1$, (c) $\beta_{uv}=1, \beta_{wv}=0$, (d) $\beta_{uv}= \beta_{wv}=0$.}
\label{image}
\end{center}
\end{figure}

The image matrix will be updated when a game is played.
The update rule is described as follows:
let $u$, $v \neq u$ and ${\cal W}$ denote the donor, the recipient and 
the set of observers respectively, chosen at random from the whole population.
The number of observers is $qN$, where $N$ is the total number of individuals in the population,
which will be assumed to be infinite for an analytical calculation.
Then the image matrix after the game is given by
\begin{eqnarray}
\beta'_{i j}=\left\{ \begin{array}{ll}
\beta_{i j} & \hspace{-10mm}(i \notin {\cal W} \vee j\neq u), \\
f^{(i)}(\alpha _{u}, \beta_{i v}) & (i \in {\cal W} \wedge j = u), \\
\end{array} \right. \label{stochas}
\end{eqnarray} 
where $\alpha_{u}$ is the action of $u$ toward $v$ in the game.
Using the abbreviation C for ``help'', D for ``refuse'',
$\alpha_{u}=\mbox{D}$ always if $\beta_{u v}=0$, or with probability $\epsilon$ if $\beta_{u v}=1$ due to the implementation error. Otherwise $\alpha_{u}=\mbox{C}$.

The function $f^{(i)}(\alpha, \beta) \in \{0,1\}$ with $\alpha \in \{\mbox{C},\mbox{D}\}$ and $\beta \in \{0,1\}$ is the assessment, from the viewpoint of $i$, if the donor chooses action $\alpha$ towards the $\beta-$recipient.
For stern-judging, $f^{(i)}(\mbox{C},1)=f^{(i)}(\mbox{D},0)=1$ and $f^{(i)}(\mbox{C},0)=f^{(i)}(\mbox{D},1)=0$ with probability $1-\mu$
 (i.e., $i$ made a correct assessment), and $f^{(i)}(\mbox{C},1)=f^{(i)}(\mbox{D},0)=0$ and $f^{(i)}(\mbox{C},0)=f^{(i)}(\mbox{D},1)=1$ with probability $\mu$  (i.e, $i$ made a mistake).

For any given updating step, only the $u-$column of the image matrix changes its values.
Since the action $\alpha_{u}$ is probabilistically determined by $\beta_{uv}$, 
the new assessment $\beta'_{w u}$ in the eyes of observer $w \in {\cal W}$ probabilistically depends on the assessments of recipient $v$ both in the eyes of $u$ ($\beta_{uv}$) and $w$ ($\beta_{wv}$) before the game.
That is to say, the updated image matrix is probabilistically determined by the old image matrix,
which yields a stochastic process on the image matrix.
This process describes the assessment dynamics, which determines who has a good image in the eyes of whom in the population.
For example, if $\beta_{u v}=1$ and $\beta_{w v}=0$,
the new assessment $\beta'_{w u}$ is 0 with probability $\bar\epsilon \bar\mu+\epsilon\mu$ and 1 with probability $\bar\epsilon\mu+\epsilon \bar \mu$ (See Fig. \ref{image} (c)).
Fig.~\ref{image} illustrates how the image matrix is updated by Eq. (\ref{stochas}) for all cases. 

\section{Results}\label{results}
\begin{figure}
\begin{center}
\includegraphics[scale=0.15]{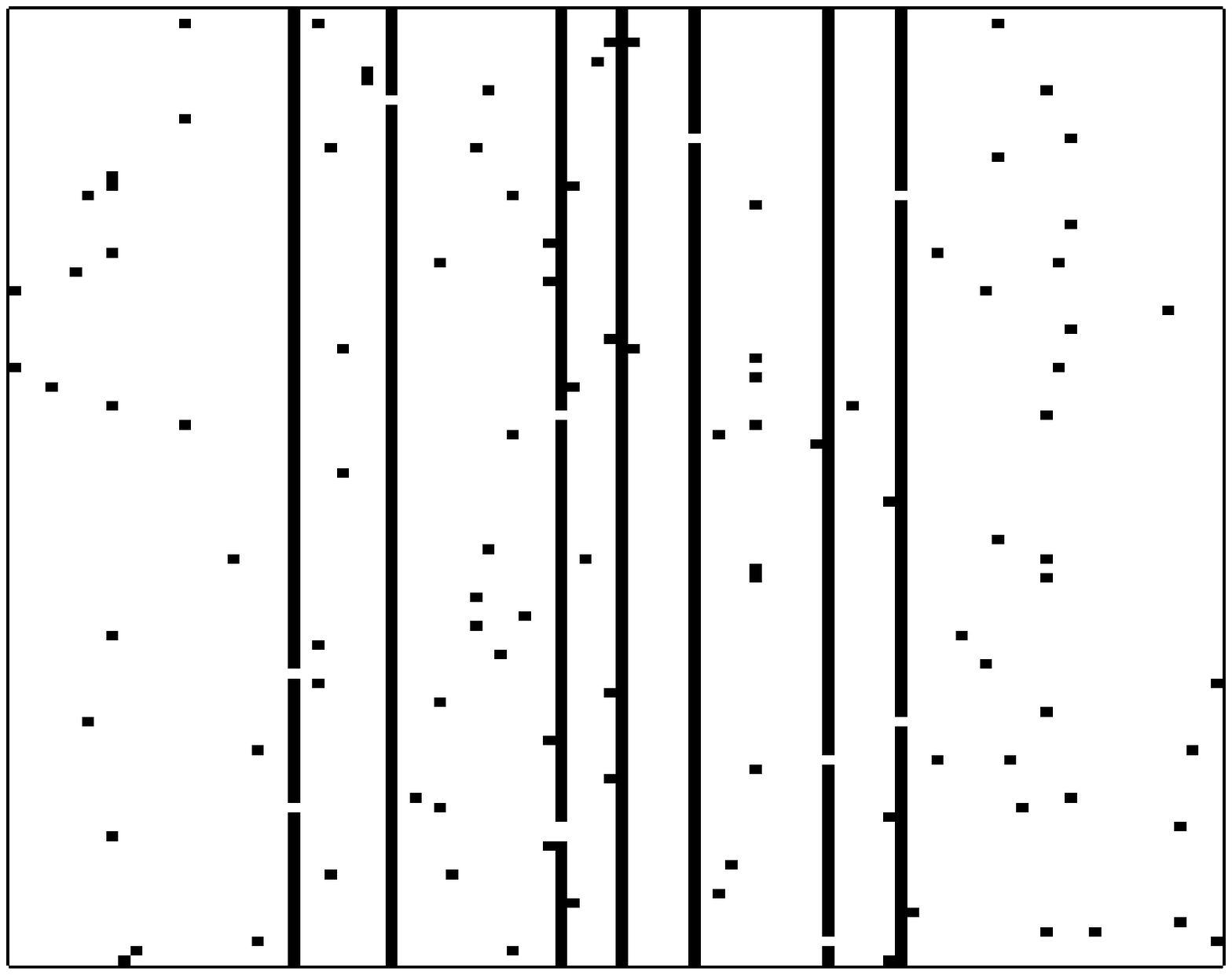}
\includegraphics[scale=0.15]{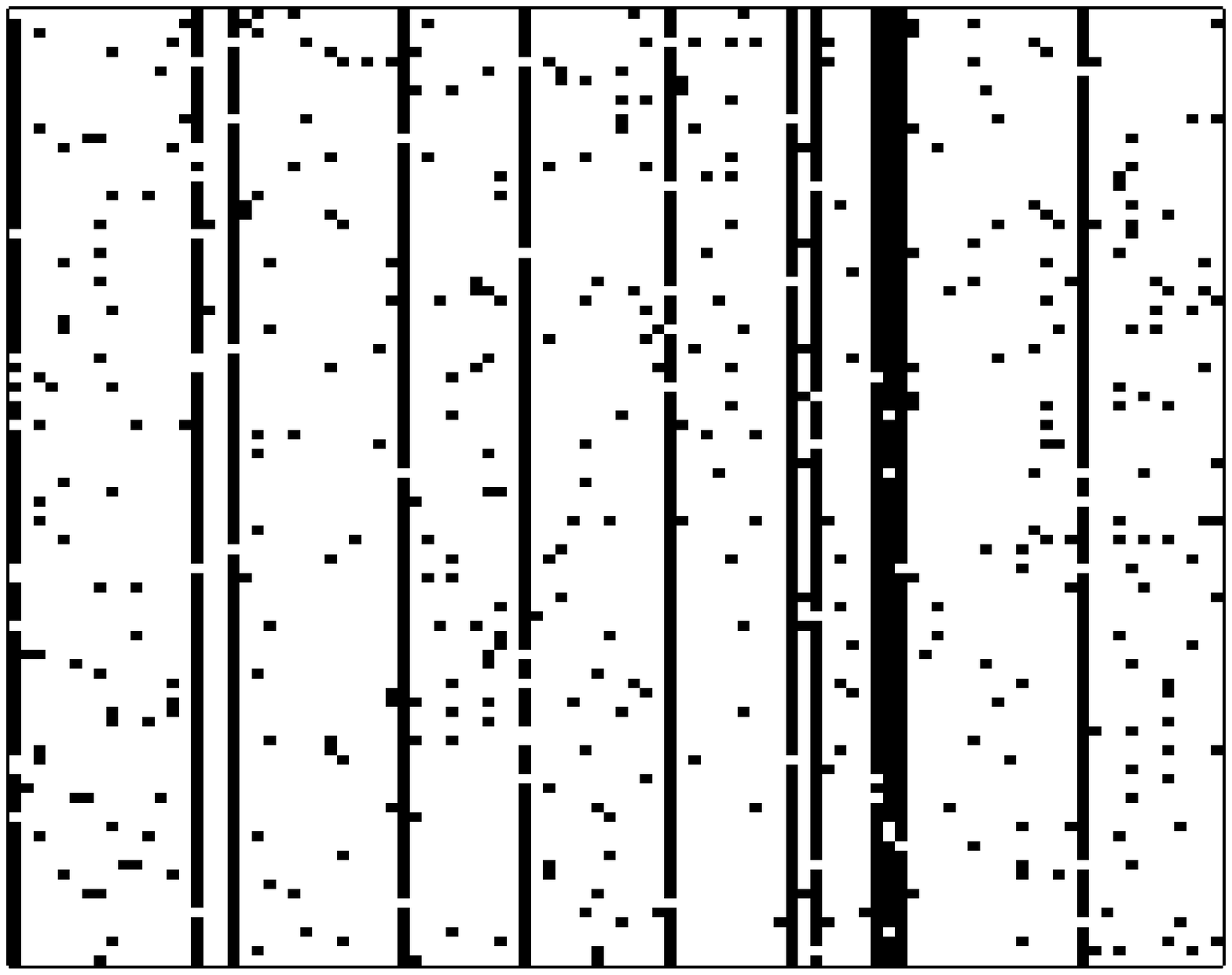}
\includegraphics[scale=0.15]{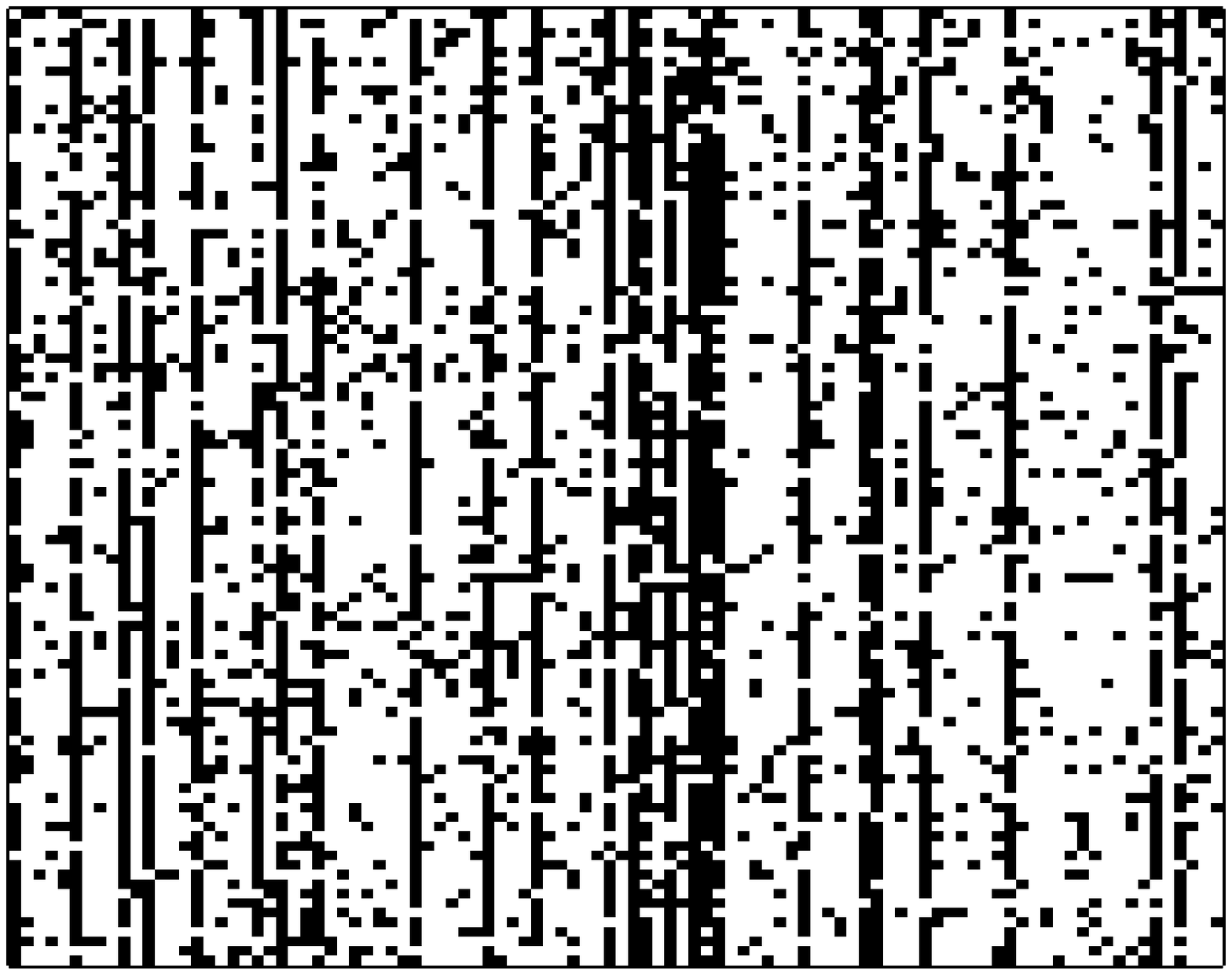}\\
\vspace{-3mm}
\hspace{2mm}{\small (50 updates)} \hspace{5mm}{\small (200 updates)} \hspace{5mm}{\small (700 updates)}\\
\vspace{4mm}
\includegraphics[scale=0.15]{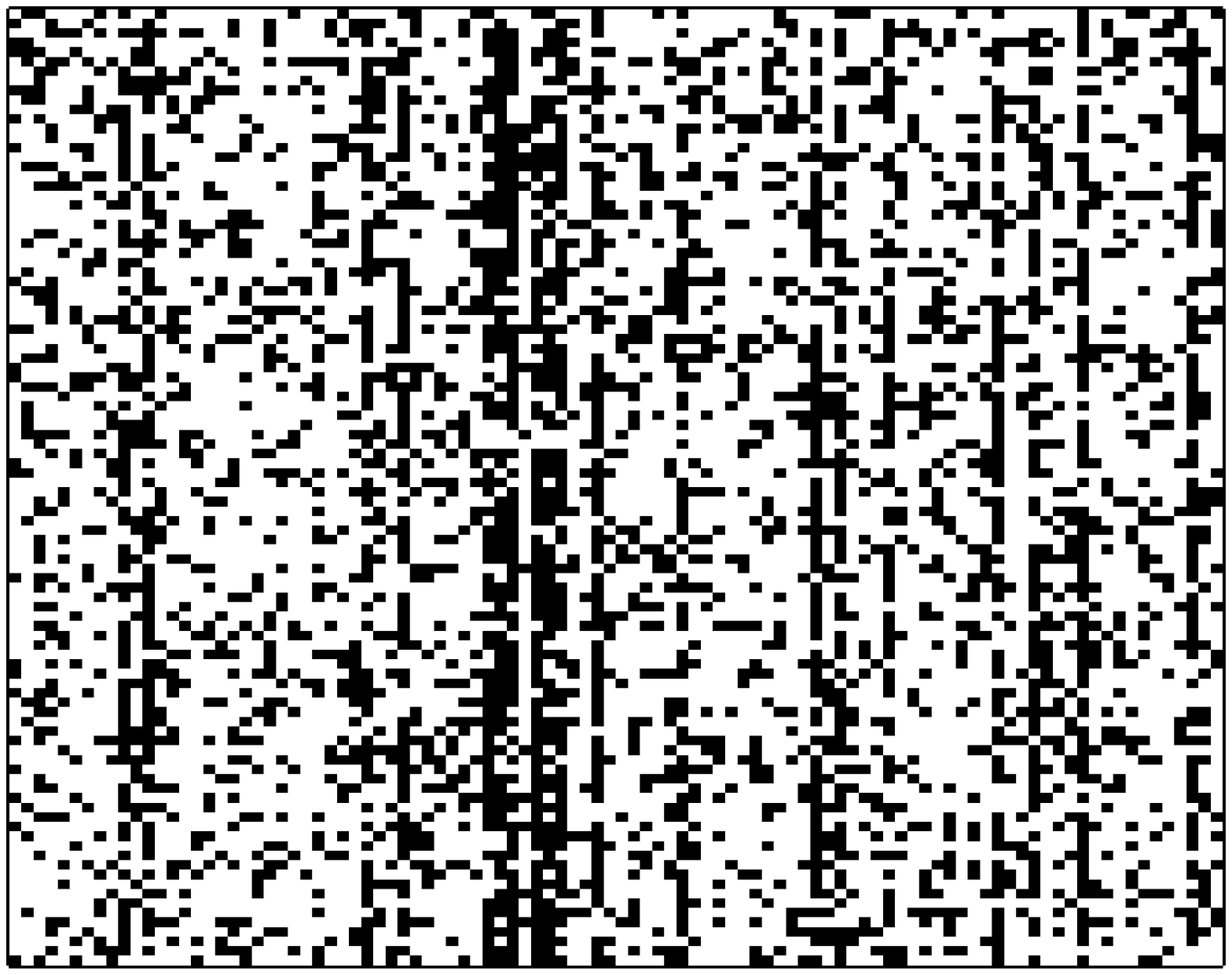}
\includegraphics[scale=0.15]{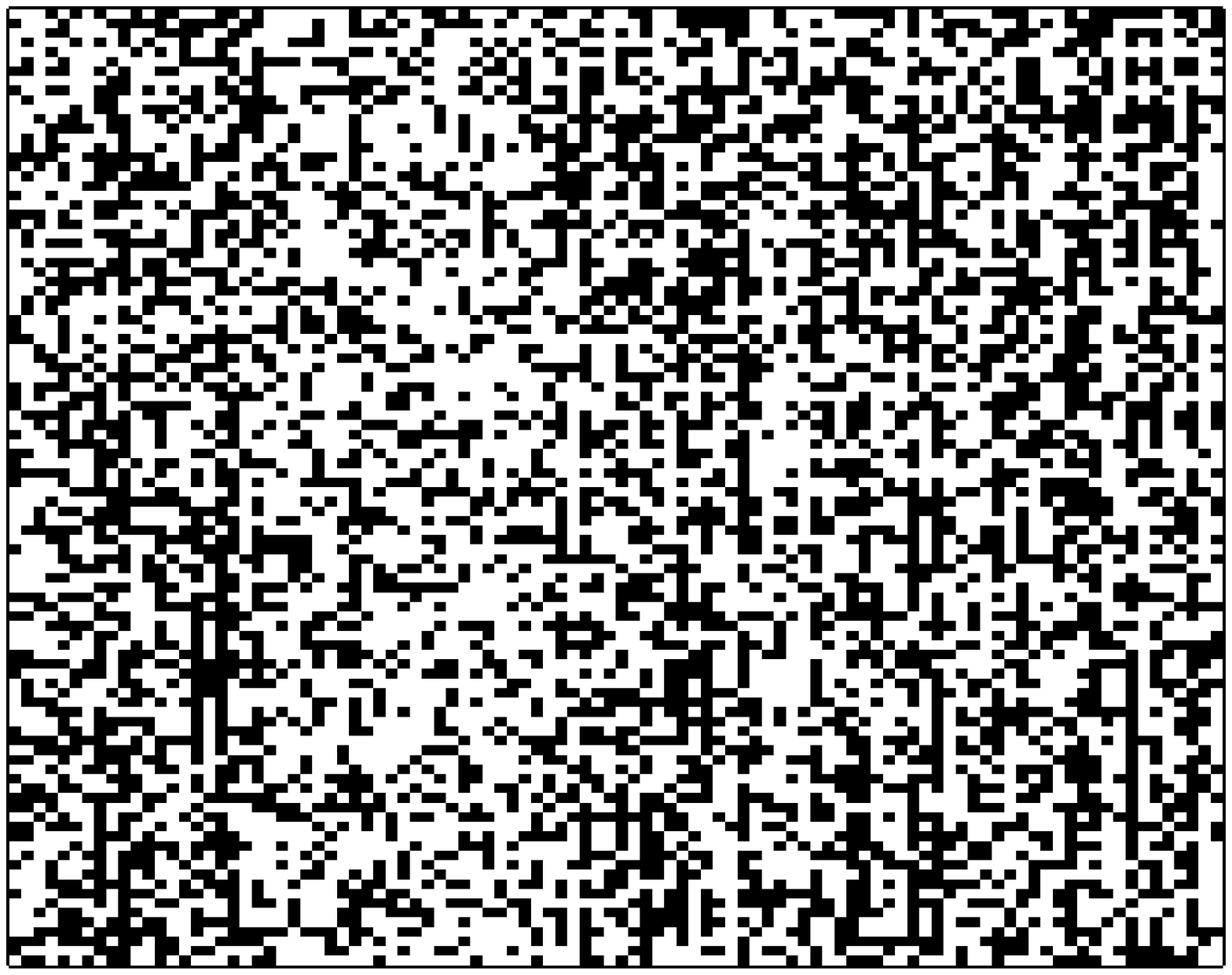}
\includegraphics[scale=0.15]{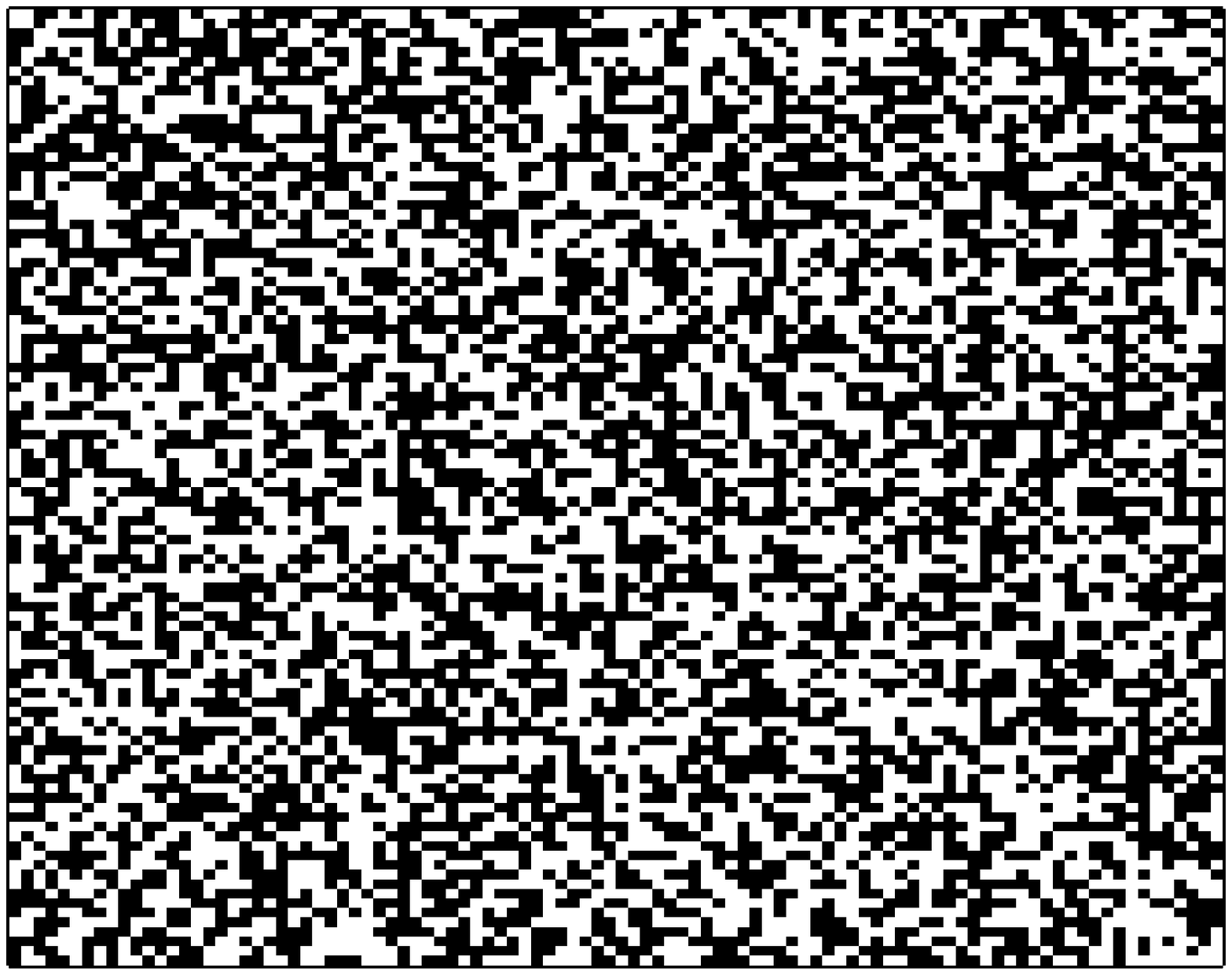}\\
\vspace{-3mm}
\hspace{0mm}{\small (1300 updates)} \hspace{3mm}{\small (2000 updates)} \hspace{3mm}{\small (5000 updates)}\\
\caption{A sample path of the time evolution of image matrix ($\beta_{ij}$) with parameters $q=0.99$ and $\mu=0.01$.
Image matrices are shown after 50 updates (top-left), 200 updates (top-middle), 700 updates (top-right),
1300 updates (bottom-left), 2000 updates (bottom-middle) and 5000 updates (bottom-right).
Initially all individuals are good:$\beta_{ij}=1$ for all $i$ and $j$.
The good image $\beta_{ij}=1$ corresponds to a white dot and the bad image $\beta_{ij}=0$ to a black dot.
Each population has 100 individuals.}
\label{pure}
\end{center}
\end{figure}

In Fig.\ref{pure}, we show an example of the time evolution of the image matrix 
with parameters $q=0.99,\mu=0.01$ and $N=100$.
We note that the diagonal elements of the image matrix  (i.e., self-images) are handled differently in simulations.
We assumed that a selected donor ($u$) can always observe the action of $u$ (i.e., $u$ is always included in the set of observers when $u$ plays).
Therefore, in each updating step, the diagonal element $\beta_{uu}$ is updated.
However, diagonal elements of the image matrix are of no interest, 
since the information of diagonal elements is never used when deciding an action in a game or when assessing others.
That is, diagonal elements do not affect off-diagonal elements, which represent images of others.

We assumed that, initially, there are only good individuals ($\beta_{ij}=1$ for all $i$ and $j$).
Due to the implementation error, bad individuals appear.
At the beginning (for instance after 50 updates), we
show some stripes in the image matrix.
This means that almost all individuals have the same image to a given player 
since $q$ is close to 1 and $\mu$ is close to $0$.
But as time evolves, mismatches between individuals spread in the image matrix
and after 5000 updates, the image matrix reaches a disordered state.
According to numerical simulations, 
the number of good individuals (i.e., the number of white dots in the image matrix) and the number of bad individuals become the same after a long period of time.

In order to derive the proportion of good individuals $r$ at equilibrium, select players $i,j$ and $k$ respectively at random from the population, and consider the situation where
$j$ is deciding whether or not to help $k$, and $i$ is a third party who can be selected as an observer.
Let us define the possible events as follows:
\begin{eqnarray}
A&=& \mbox{ $i$ is selected as an observer},  \nonumber \\
B&=& \mbox{ $j$ intends to help $k$ and help is actually given} ,\nonumber \\
C &=& \mbox{ $i$ positively assesses the action of $j$ in event $B$}, \nonumber \\
D &=& \mbox{  $j$ intends to help $k$ but help is not given} ,\nonumber \\
E &=& \mbox{ $i$ positively assesses the action of $j$ in event $D$}, \nonumber \\
F &=& \mbox{ $j$ decides not to help $k$},\nonumber \\
G &=& \mbox{ $i$ positively assesses the action of $j$ in event $F$}, \nonumber \\
H &=& \mbox{ $i$ is not selected as an observer}, \nonumber \\
I &=& \mbox{ $i$ has a good image of $j$}.\nonumber
\end{eqnarray}

Then at equiibrium, the following equality holds:
\begin{equation} 
 r = P[A](P[B\wedge C] +P[D \wedge E]+ P[F \wedge G])+P[H \wedge I],
 \label{req}
\end{equation}
where $P[X]$ is the probability that event $X$ occurs.
For example, when the events $A,B$ and $C$ occur at once,
a randomly chosen individual $i$ has a good image of another randomly chosen individual $j$ after the game.

The probabilities in Eq.(\ref{req}) are given by
\begin{eqnarray}
P[A]&=& q, \nonumber \\
P[B \wedge C]&=& \bar\epsilon (r_{11}\bar\mu+r_{10}\mu ), \nonumber \\
P[D \wedge E]&=& \epsilon (r_{10}\bar\mu+r_{11}\mu ), \nonumber \\
P[F \wedge G]&=& (r_{00}\bar\mu+r_{01}\mu ), \nonumber \\
P[H \wedge I] &=& (1-q) r, \nonumber
\end{eqnarray}
where $r_{11}$ is the probability that two random players ($i,j$) both have a good image of another random player ($k$),
$r_{10}$ is the probability that a random individual $j$ as a donor has a good image of $k$ but a random individual $i$ as an observer
has a bad image of $k$, and so on.
Since $r_{10}=r_{01}$ holds because we select $i$ and $j$ at random,
these probabilities can be denoted by a common symbol $r_{1/2}$.

For the case of perfect information ($q=1$) without an assessment error ($\mu=0$),
$r_{11}=r, r_{1/2}=0$ and $r_{00}=1-r$ because there is no difference in the opinions of any two individuals.
In this case, Eq.(\ref{req}) yields the solution $1/(1+\epsilon)$ \cite{Ohtsuki06,Ohtsuki07}.
For the opposite extreme case where $q$ is very small,
the mean field approximation is valid: $r_{11}=r^2, r_{1/2}=r(1-r)$ and $r_{00}=(1-r)^2$ \cite{Uchida10b,Sigmund12}.
Then from the Eq.(\ref{req}), we find two solutions:
one is $r=1/2$ regardless of $\epsilon$ and $\mu$ and the other is the meaningless solution $r=\bar\mu/(\bar\epsilon(1-2\mu))>1$.

For the intermediate cases where $q<1$ or  $\mu>0$,
we need additional three equations to estimate the second order unknowns $r_{11},r_{1/2}$ and $r_{00}$.
For two of these three equations, we can use trivial relations between these variables:
$r_{11}+r_{10}=r$ and $r_{10}+r_{00}=1-r$. Therefore, the essential second order unknown is only $r_{11}$.

Now let us randomly pick two individuals $i_1$ and $i_2$ who may observe the action of $j$.
We define the possible events as follows:
\begin{eqnarray}
A_2&=& \mbox{ $i_1$ and $i_2$ are selected as observers},  \nonumber \\
C_2 &=& \mbox{ $i_1$ and $i_2$ positively assess the action of $j$ in $B$}, \nonumber \\
E_2 &=& \mbox{ $i_1$ and $i_2$ positively assess the action of $j$ in $D$}, \nonumber \\
G_2 &=& \mbox{ $i_1$ and $i_2$ positively assess the action of $j$ in $F$}. \nonumber
\end{eqnarray}

Then we find the equality for $r_{11}$:
\begin{eqnarray} 
 r_{11} =P[A_2] (P[B\wedge C_2] +P[D \wedge E_2] + P[ F \wedge G_2] )\nonumber \\
+ 2 P[A]  (P[B\wedge C] +P[D \wedge E] + P[F \wedge G]) \nonumber \\
\times P[H \wedge I] +(1-q)^2 r_{11}. \label{req2}
\end{eqnarray}
For instance, $P[A]P[B\wedge C] $ in the second term is the probability
 that $i_1$ is an observer who perceives the action of $j$ as good after $j$ helps
and $P[H \wedge I] $ is the probability that $i_2$ is not an observer who has a good image of $j$.
If the events $A,B,C,H$ and $I$ occur, both $i_1$ and $i_2$ have a good image of  $j$ after the game.

Note that these probabilities are independent, since the probability 
$P[A]P[B\wedge C] $ depends on the current image of $k$ in the eyes of $i_1$ and $j$,
 wheres $P[H \wedge I] $ is a function of the current image of $j$ in the eyes of $i_2$
 (because the assessment of $j$ in the eyes of $i_1$ and $j$ and that of $k$ in the eyes of $i_2$ have been made independently).

The last them $(1-q)^2 r_{11}$ is the probability that $i_1$ and $i_2$ who have a good image of $j$ are not selected as observers.
The probabilities in the first term are given by
\begin{eqnarray}
P[A_2]&=& q^2, \nonumber \\
P[B \wedge C_2]&=& \bar\epsilon (r_{111}\bar\mu^2 + (r_{101}+r_{110})\bar\mu\mu + r_{100}\mu^2 ), \nonumber \\
P[D \wedge E_2]&=& \epsilon (r_{111}\mu^2 + (r_{101}+r_{110})\bar\mu\mu + r_{100}\bar\mu^2 ), \nonumber \\
P[F \wedge G_2]&=& r_{000}\bar\mu^2 + (r_{001}+r_{010})\bar\mu\mu + r_{011}\mu^2, \nonumber
\end{eqnarray}
where, for example, $r_{111}$ is the probability that $j,i_1$ and $i_2$ all have a good image of $k$
and $r_{101}$ that $j$ and $i_2$ have a good image but $i_1$ has a bad image of $k$; and so on.
Again $r_{110}=r_{101}=r_{011}=:r_{2/3}$ and $r_{001}=r_{010}=r_{100}=:r_{1/3}$ hold.
Therefore we have four third order unknowns ($r_{111},r_{2/3},r_{1/3}$ and $r_{000}$).
Furthermore because of the trivial relations $r_{111}+r_{2/3}=r_{11}, r_{2/3}+r_{1/3}=r_{1/2}, r_{1/3}+r_{000}=r_{00}$,
the essential third order unknown is only $r_{111}$.

However, taking into account all of the obtained relations above and calculating the first term in Eq.(\ref{req2}),
we find that the third order variable is canceled out:
\begin{eqnarray}
P[B\wedge C_2] +P[D \wedge E_2] + P[F \wedge G_2] =(\bar\epsilon \bar\mu^2+\epsilon \mu^2) r_{11}\nonumber \\
+(\bar\epsilon \mu^2-\bar\epsilon \bar\mu^2+2\bar\mu\mu)r_{1/2}+\bar\mu^2 r_{00}. \nonumber
\end{eqnarray}

Thus Eqs.(\ref{req}) and (\ref{req2}) yield a closed equation system for unknowns $r$ and $r_{11}$.
Especially, Eq.(\ref{req})  is linear in $r$ and $r_{11}$ and we can express $r_{11}$ by $r$.
Inserting this expression into $r_{11}$ in Eq.(\ref{req2}), we find a quadratic equation for $r$.
Solving this equation leads to two solutions,
namely $r=1/2$ and the meaningless solution $r=R_1/R_2$ with $R_1=\bar\mu(1-q+\mu q+\epsilon \mu q-2 \epsilon \mu^2 q)>R_2=\bar \epsilon (1-2\mu)(1-q)$.
That is to say, $r=1/2$ is the unique solution for $q<1$ or for $\mu>0$.

\section{Discussion}
We found that private information and assessment errors have a striking effect on stern-judging:
Any difference or mismatch between images among individuals due to private information
 and assessment errors spread in a population
so that individuals assess each other as good at random with probability 1/2 in the end.

But why does stern-judging collapse, even though it is highly successful in the case of public information without personal assessment errors?
The point is that stern-judging does not have any mechanism (or a function) which repairs a mismatch in the opinions  between two observers.
In fact,  if two observers with stern-judging disagree about a given player, 
this difference remains, no matter which action the observed player chooses and whoever the recipient is.
Therefore once a new mismatch between two individuals has appeared, 
these individuals continue to have different opinions. 
As a result, the image matrix becomes totally disordered.

This indicates that, besides private information and personal assessment errors,
there can be more factors that cause mismatches
 and lead to a collapse of stern-judging.
 For instance, in this paper, the imperfectness of information can be interpreted both by direct and indirect observation models.
In our model, players in a population can obtain information about the interactions of others either by direct observation, or by rumor or gossip.
In the case of rumor or gossip,  the transmitted information is assumed to always be correct
 even though the information is not passed to all individuals.
In the literature however, there is a model that includes incorrect rumors \cite{Nakamaru04}. 
Such incorrect rumors can also cause a mismatch of images if the rumor is transmitted individually,
therefore they can also lead to the collapse of stern-judging.

Note that our definition of imperfectness of information is just one form.
Other forms of imperfect information can be found in literature \cite{Brandt05,Brandt06,Mohtashemi03,Nowak98b}.
Nowak and Sigmund \cite{Nowak98b} assumed that each player either knows or does not know the reputation of another player, and assumes a good reputation by default (if a player does not know the reputation).
Further, Brandt and Sigmund \cite{Brandt06} allowed $q$ to grow with age and found that cooperation can then be stably maintained.
Mohtashemi and Mui \cite{Mohtashemi03} modified the model of Nowak and Sigmund \cite{Nowak98b} 
by assuming that the information spreads through social networks. 
They derived a condition for $q$ which is similar to the inequality found by Nowak and Sigmund \cite{Nowak98b}.
In those studies, the information is not perfect but public.
How this imperfectness affects stern-judging is an interesting question,
since, in this case, a mismatch between individuals can be repaired because the information is public.

In this paper, the effects of private information and assessment errors on other best-studied assessment models such as scoring (first-order) \cite{Nowak98a} or standing (second-order) \cite{Sugden86} are not investigated.
Standing is the other second-order assessment rule in the leading eight.
Since there is experimental evidence to support the view that higher-order assessment can overtax human cognitive abilities \cite{Milinski01}, 
it will be interesting to theoretically investigate the effects of factors that cause disagreements between individuals on those assessment rules.
At least in this paper, we showed that private information and assessment errors strikingly affect one of the best-known assessment models.

\section*{Acknowledgments}
Satoshi Uchida wishes to thank Toshihiro Shimizu at Kokushikan University for his useful comments.
Part of this work was supported by grant RFP-12-21 from the Foundational Questions in Evolutionary Biology Fund.

\bibliographystyle{elsarticle-num}
\bibliography{ref}

\end{document}